\documentclass[fleqn,useAMS,usenatbib]{mn2e}
\pdfoutput=0

\usepackage{longtable,lscape}
\usepackage{amsmath}
\usepackage{graphicx}
\usepackage{aas_macros}
\usepackage{amssymb}
\usepackage{calc,float,amsmath,booktabs}
\usepackage{fixltx2e} 

\providecommand		{\adsurl}[1]		{\href{#1}{ADS}}

\newcommand       \be        {\begin{equation}}
\newcommand       \ee        {\end{equation}}
\newcommand       \kms		{\,{\rm km \,\, s}^{-1}}
\newcommand       \cms		{\,{\rm cm \,\, s}^{-1}}
\newcommand       \mms		{\,{\rm mm \,\, s}^{-1}}
\newcommand       \cm		{\,{\rm cm }}

\newcommand       \rsun		{\,{\rm R_\odot}}

\newcommand       \yr		{\,{\rm yr }}

\newcommand       \K			{\,{\rm K }}

\newcommand       \muHz		{\,\mu{\rm Hz }}
\newcommand       \mumag		{\,\mu{\rm mag }}
\newcommand       \perday	{\,{\rm d}^{-1}}
\newcommand       \erg		{\,{\rm erg }}

\newcommand       \Lc		{\rm L_{conv}}
\newcommand       \Emode		{\rm E_{mode}}
\newcommand       \Nm		{\rm N_{m}}

\newcommand       \Mc        {{\cal M}_{c}}

\newcommand       \Edotg		{\dot{E}_{g}}
\newcommand       \grad		{\rm \gamma_{rad}}

\newcommand       \vg		{v_{\rm group}}

\newcommand       \omc		{\omega_{c}}

\newcommand       \om		{\omega}
\newcommand       \tauw		{\tau_{w}}
\newcommand       \vdisk		{v_{\rm disk}}

\newcommand       \rmsdm		{\delta m_{\rm rms}}

\newcommand       \nuc		{\nu_{c}}

\newcommand       \brunt		{Brunt-V\"ais\"al\"a\ }
\newcommand 		  \lmax		{\ell_{\rm max}}

\def \Msun{M$_\odot$}

\title[Convectively Excited Gravity Waves]
	{The Observational Signatures of Convectively Excited Gravity Modes in Main Sequence Stars}
	
\author[Shiode, et~al.]{Joshua~H.~Shiode,$^{1}$\thanks{E-mail: jshiode@astro.berkeley.edu} Eliot~Quataert,$^{1}$
	Matteo Cantiello,$^{2}$ and Lars Bildsten$^{2}$ \\
$^{1}$Department of Astronomy \& Theoretical Astrophysics Center, University of California, Berkeley, CA 94720-3411, USA\\
$^{2}$Kavli Institute for Theoretical Physics, Kohn Hall, University of California, Santa Barbara, CA, 93106, USA\\
}

\begin{document}
\maketitle
\label{firstpage}

\begin{abstract}

We predict the flux and surface velocity perturbations produced by convectively excited gravity modes (g-modes) in main sequence stars. Core convection in massive stars can excite g-modes to sufficient amplitudes to be detectable with high precision photometry by \textit{Kepler} and CoRoT,  if the thickness of the convective overshoot region is $\lesssim 30$ per cent of a pressure scale height.  The g-modes manifest as excess photometric variability, with amplitudes of $\sim$ 10 micromagnitudes at frequencies $\lesssim 10 \muHz$ ($0.8 \perday$) near the solar metallicity zero-age main sequence.  The flux variations are largest for stars with $M \gtrsim 5$ \Msun{}, but are potentially detectable down to $M \sim 2 - 3$ \Msun{}. During the main sequence evolution, radiative damping decreases such that ever lower frequency modes reach the stellar surface and flux perturbations reach up to $\sim$ 100 micromagnitudes at the terminal-age main sequence. Using the same convective excitation model, we confirm previous predictions that solar g-modes produce surface velocity perturbations of $\lesssim 0.3 \mms$.  This implies that stochastically excited g-modes are more easily detectable in the photometry of massive main sequence stars than in the Sun.

\end{abstract}

\begin{keywords}
{stars: oscillations --- stars: interiors --- convection} 
\end{keywords}


\section{Introduction} \label{sec:intro}

At the interfaces between convective and radiative zones in stellar interiors, convective motions transfer a fraction of their kinetic energy into waves in the radiative layer \citep[e.g.,][]{press1981,goldreich1990,belkacem2008}. Studying the details of this energy transfer is important for our understanding of mixing at convective boundaries, the evolution of shear layers and the excitation of stellar oscillations in many contexts. 

The Sun provides an exquisite laboratory for studying convectively excited sound waves (p-modes), with a forest of modes observed at the solar surface \citep[see][for recent reviews]{gizon2010,dalsgaard2002}. These sound waves have been detected in many other stars as well, with rapid growth in the sample of observed ``solar-like oscillators'' in the past several years thanks to high precision monitoring campaigns like \textit{Kepler} \citep{bedding2011}. 

To date, gravity modes (g-modes) have not been convincingly observed at the solar surface \citep[see][for a recent review]{appourchaux2010}. Gravity modes are observed in several other stellar types, including, for example, white dwarfs \citep{winget2008} and slowly-pulsating B-stars \citep{de-cat2007}. In each of these cases, the g-modes are linearly unstable, rather than stochastically excited as in the case of solar-like oscillations. In the interesting case of the mixed gravity and pressure modes recently observed in giants \citep[see e.g.,][]{bedding2011a,beck2011sci}, the convective excitation occurs in the envelope where the modes behave locally as pressure waves. Thus, the \emph{direct} excitation of g-modes by turbulent convection is not well tested observationally.




Main sequence stars more massive than the Sun have convective cores and relatively compact, predominantly radiative envelopes, through which g-modes propagate. This structure makes them potential hosts for observable, stochastically excited g-modes. The unprecedented micromagnitude precision of photometric monitoring campaigns like \textit{Kepler} \citep{kepler} and CoRoT \citep[Convection, Rotation and planetary Transits;][]{corot} provides exciting prospects for detecting these stochastically excited g-modes and opening a new window into both the physics of convective boundaries and massive stellar interiors. Indeed, \citet{samadi2010} studied the convective excitation of g-modes in main sequence stars with masses of 10, 15, and 20 \Msun{} and central hydrogen mass fraction of 0.5, finding that convectively excited g-modes in these stars may reach amplitudes near the threshold for detectability with CoRoT. In the following, we present a complementary approach and investigate g-mode amplitudes in a wider range of initial stellar masses, from 2 to 30 \Msun{}, and evolutionary states along the main sequence. 


We begin by describing our model for the spectrum of g-mode excitation in \S \ref{sec:theory}, followed by our method for calculating the observable signatures of convectively excited g-modes in \S \ref{sec:mods}, \ref{sec:waves}, and \ref{sec:quasiad}. In \S \ref{sec:res}, we present the results of our calculations for the surface flux and velocity perturbations of g-modes in massive main sequence stars. We conclude with a discussion of remaining uncertainties, including the effects of rotation and stellar evolution modeling, and the current and ongoing observations which might shed light on these predictions (\S \ref{sec:disc}).


\section{Convective Excitation of Gravity Modes} \label{sec:theory}

Convection efficiently excites gravity modes in an adjacent stably stratified medium, as both convective and g-mode perturbations are roughly incompressible \citep{goldreich1990}.  The power spectrum of g-modes excited by stellar convection at and above the characteristic convective turnover frequency, $\omc$, has been calculated using both heuristic physical arguments (setting the wave pressure in the stable layer equal to the convective ram pressure; \citealt{press1981, spruit1991}) and by solving the inhomogeneous wave equation with convective source terms \citep{press1981,goldreich1990,kumar1999,belkacem2009}.   These approaches are reviewed and extended to more realistic radiative-convective boundaries in \citet{lecoanet2012}; we summarize here the key results for our application.

The low frequency g-modes that carry most of the wave power are strongly damped by radiative diffusion (see \S \ref{sec:waves}).   As a result only g-modes with frequencies significantly larger than the characteristic convective turnover frequency (i.e., $\omega \gg \omega_c$) can set up global standing waves potentially detectable at the stellar surface -- this is true for both solar-type stars with envelope convection zones and massive stars with core convection zones.   High frequency g-modes have characteristic radial wavelengths in the radiative zone that are large compared to the thickness of the radiative-convective boundary.  In this limit, it is reasonable to approximate the radiative-convective boundary as a discontinuity in the \brunt frequency (as in \citealt{goldreich1990}).   For g-modes excited by a core convection zone (or its overshoot region), which is relatively unstratified, the excitation  occurs at a roughly fixed position at the edge of the convection zone.   The power spectrum of energy supplied to g-modes is then given by
\be \label{eqn:spect}
	\begin{split}
		\frac{d \Edotg}{d \ln \om \, d \ell} \sim & \: \Mc \, \Lc  \left(\frac{\om}{\omc}\right)^{-a} \\
			& \:\: \ell^{b} \, \left(\frac{H}{r}\right)^{b+1} \left(1 + \ell \frac{H}{r}\right) \,
			 \exp \left[-\left(\frac{\ell}{\lmax}\right)^{2}\right]
	\end{split}
\ee
over the range
\be
\om \geq \omc \equiv v_{\rm conv}/H, \;  \: \ell \geq 1, \label{eqn:speclims}
\ee
where $\Lc$ is the total luminosity carried by convection, $\Mc \left(\equiv v_{\rm conv} / c_{\rm sound}\right)$ is the convective Mach number,  $v_{\rm conv}$ is the convective velocity, $\lmax \equiv (r / H) \, (\om / \om_{c})^{3/2}$, and $r$ and $H$ are evaluated at the edge of the convective core (where their ratio is $\sim 1$). The power-law exponents $a$ and $b$  are given by $a = 13/2$ and $b = 2$.\footnote{Because excitation of g-modes by Reynold's stresses dominate excitation by entropy fluctuations \citep{goldreich1990}, the convective luminosity in eqn. \ref{eqn:spect} is technically that associated with the kinetic energy, rather than enthalpy.  In mixing length theory, these two contributions to the convective luminosity are comparable.     For the purposes of our calculation, we thus take $\Lc$ to be the total convective luminosity in the stellar model and absorb uncertainties in the relative contribution of kinetic and enthalpy fluxes into the uncertain normalization of eqn \ref{eqn:spect}.}

Note that integrating over frequency and $\ell$, equation \ref{eqn:spect} implies that convection excites a total g-mode luminosity of \citep{goldreich1990}
\be
	\Edotg \sim \Mc \, \Lc. \label{eqn:lwave}
\ee
In fact, the  efficiency of g-mode excitation can exceed equation \ref{eqn:lwave} depending on the nature of the convective-radiative transition \citep{press1981,lecoanet2012}.   Equation \ref{eqn:spect} is, however, appropriate for the high frequency g-modes of interest in this paper.  In our numerical work, we choose the uncertain dimensionless constant in equation \ref{eqn:spect} so that the integrated wave power is exactly $\Edotg = \Mc \, \Lc$.   

One of the primary uncertainties in the prediction of g-mode excitation is whether the excitation occurs primarily in the bulk of the convection zone or in a thin ``overshoot" region near the radiative-convective boundary.   We parameterize this uncertainty by taking the local scale-height in the excitation region to be $\eta H$, where $H$ is the true pressure scale height near the radiative-convective boundary.   This amounts to taking $H \rightarrow \eta H$ in equation \ref{eqn:spect} when calculating the wave excitation.   The parameter $\eta$ can take on values in the range $(0, 1]$, such that $\eta = 1$ corresponds to excitation dominated by eddies of size $\sim H$ in the convection zone and $\eta < 1$ to excitation dominated by a thin overshoot region hosting eddies of size $\sim \eta H$. Excitation dominated by a thin overshoot region will have a higher characteristic frequency $\omc / \eta$ since the length scale of the convective motions is reduced by $\eta$. This serves to shift  the g-mode power input to higher frequency and shorter spatial scales (higher $\om$ and $\ell$).  Note, in particular, that for low $\ell$ modes having $\omega \gg \omega_c$, $\dot E_g \propto H^{-7/2}$.  Thus excitation in a thin overshoot layer significantly increases the power supplied to the high frequency modes that are the most observable.   

In their numerical simulations of the solar radiative-convective transition, \citet{rogers2006} found that the power into  g-modes is roughly constant for $\omega \sim 1-10 \, \omega_c$ and then decreases significantly for $\omega \gtrsim 10 \, \omega_c$ (see their figs. 1 \& 2).   This increase in the characteristic frequency of the excited g-modes is reasonably consistent with excitation in the overshoot layer given the width of the overshoot layer of $\sim 0.05$ H found in the same simulations (see \citealt{rogers2006a}).  The smaller scales in the overshoot region would also naturally produce power at higher $\ell$, as also found by \citet{rogers2006}.\footnote{Note that the simulations of \citet{rogers2006} use an artificially high convective luminosity relative to the solar value.  It is unclear whether this changes the functional form of the g-mode power vs. frequency;  this needs to be studied in more detail.}

\begin{figure*}
\centering
\includegraphics[width=\textwidth]{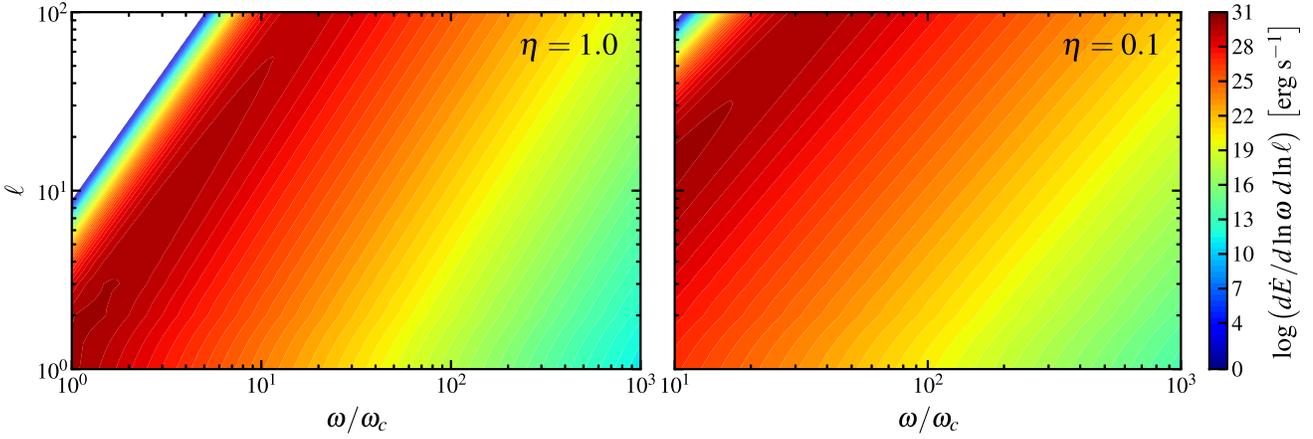}
\caption[g-mode excitation spectrum]{Convective excitation spectrum vs. $\om / \omc$ (abscissa) and $\ell$ (ordinate), for g-modes in our 2 \Msun{} model. The spectrum shown in the left panel is for excitation by the convection zone, $\eta = 1$, and the right for excitation dominated by an overshoot region having a width of 10 per cent of a pressure scale height, $\eta = 0.1$. Note the different scales on the abscissa in the two panels, since our parametrization of the power spectrum applies only above $10 \, \omc$ for $\eta = 0.1$ (see eqn. \ref{eqn:speclims}). Also note values for $\log d\Edotg / (d \ln \om \, d\ln \ell) < 0$ are represented as white. The spectra for models with $M \gtrsim 10$ \Msun{} peak at higher $\ell$ values since $r / H \gtrsim 2$ at the edges of their convective cores; otherwise, spectra for different masses are qualitatively similar, with the overall energy injection rate scaled according to the value of $\Mc \, \Lc$ (see Table \ref{tab:conv}).} \label{fig:dLex}
\end{figure*}

Fig. \ref{fig:dLex} shows the logarithmic excitation spectrum for our 2 \Msun{} model, for $\eta = 1$ (left panel) and $0.1$ (right panel). This shows that most of the wave energy is in the lowest frequency modes convection can excite. In addition, since the wavelength and frequency of convective eddies are correlated, energy input at higher mode frequencies is predominantly in modes of higher $\ell$. 


\section{Stellar Models} \label{sec:mods}

We have constructed main sequence models from zero-age main sequence (ZAMS) to the terminal-age main sequence (TAMS), at a central hydrogen mass fraction of 0.01, for a range of initial masses from 2 to 30 \Msun{} using the MESA stellar evolution code \citep[version 4298;][]{mesa2011}\footnote{http://mesa.sourceforge.net/}. All models are non-rotating and solar metallicity (which we take to be $Z = 0.02$) with the \citet[]{gs98} chemical mixture. Following the results of \citet{brott2011}, we determine convective boundaries using the Ledoux criterion with 30 per cent of a pressure scale height of overshoot,\footnote{We use a step function overshoot prescription, in which the convection zone is extended a distance of 30 per cent of a pressure scale height above the Ledoux boundary, with a constant diffusion coefficient.} and a mixing length parameter $\alpha = 1.5$. We also use semiconvection with a dimensionless efficiency parameter, $\alpha_{sc} = 0.1$, though this is largely irrelevant due to the overshoot. Finally, we assume the theoretical mass loss rates of \citet*{vink2001} (or \citealt{de-jager1988} when $T_{\rm eff} < 10^{4} \K$) scaled down by a factor of 0.8.

\begin{figure}
\centering
\includegraphics[width=\columnwidth]{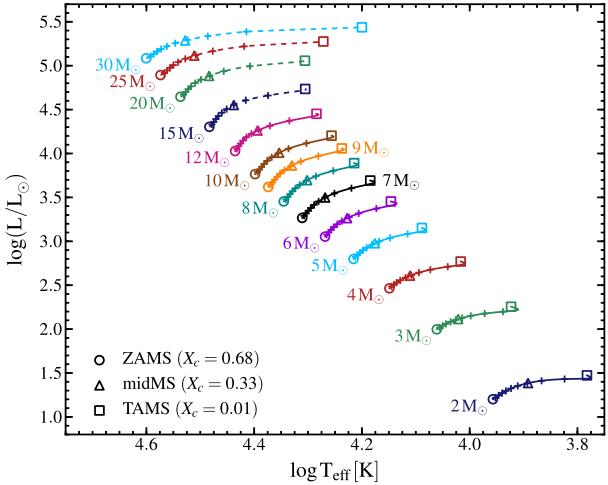}
\caption[H-R Diagram]{H-R Diagram showing the evolutionary tracks of our 2 to 30 \Msun{} model sequences spanning ZAMS to TAMS ($X_{c} = 0.01$). The circles, triangles and squares mark the locations of models for which we have calculated convectively excited modes, and are in the evolutionary states given in the legend. Crosses mark every 1/10th of the main sequence lifetime (i.e., the model age at TAMS).}\label{fig:hrd}
\end{figure}

We tested for numerical convergence by varying the number of mesh points and timesteps taken during the evolution, finding good agreement at the level of refinement used for the models presented.\footnote{\texttt{mesh\_delta\_coeff} = 0.25 and \texttt{varcontrol\_target} = $10^{-4}$} We have also tested a range of other model parameters, including the mixing length, boundary definitions and rotation, to test their effects on our results (see \S \ref{ssec:StEvo}). Fig. \ref{fig:hrd} shows the evolutionary tracks in the H-R diagram for the models described above from ZAMS to TAMS. Along the sequence, crosses appear every 1/10th of the main sequence lifetime, and symbols mark the locations where we have calculated the convective excitation of g-modes: ZAMS ($X_{c} = 0.68$), midMS ($X_{c} = 0.33$), and TAMS ($X_{c} = 0.01$).


Table \ref{tab:conv} gives parameters for the convective cores of our ZAMS models, including the outer scale convective turnover frequency ($\nuc \equiv \omc / (2\pi)$), the core luminosity ($\Lc$, which is roughly equivalent to the emergent stellar luminosity) and the convective mach number ($\Mc$). Over the range of masses represented, $\Mc$ increases with luminosity (and mass), but only as $\sim \Lc^{0.2}$. The yet weaker dependence of the convective turnover frequency on mass reflects the increase in convective core radius with mass, since energy-bearing eddies span roughly the core radius.


\begin{table}
	\label{tab:conv}
	\centering
	\begin{minipage}{\columnwidth}
		\caption{Core Convection Parameters}
    		\begin{tabular}{@{}ccccccc@{}}
      		\toprule
       		Mass\footnote{Initial stellar mass} & 
       			$\Lc$\footnote{Convective luminosity on ZAMS} & 
				$\Mc$\footnote{Convective mach number on ZAMS} & 
				$\nuc$\footnote{Convective turnover frequency on ZAMS} & 
       			\multicolumn{2}{c}{$\nu_{\rm min} \, \left[\mu{\rm Hz}\right]$\footnote{
					Minimum frequency for $\ell = 1$ standing waves}}\\ 
			$\left[{\rm M}_{\odot}\right]$ & $\left[{\rm L}_\odot\right]$ & 
				& $\left[\mu{\rm Hz}\right]$ & ZAMS & TAMS\\
      		\midrule
			$2$ & $14$ & $4.37 \times 10^{-5}$ & $0.05$ & $5.81$ & $1.58$\\
			$3$ & $75$ & $6.67 \times 10^{-5}$ & $0.07$ & $5.91$ & $1.74$\\
			$4$ & $228$ & $8.75 \times 10^{-5}$ & $0.08$ & $5.54$ & $1.71$\\
			$5$ & $579$ & $1.23 \times 10^{-4}$ & $0.10$ & $4.73$ & $1.65$\\
			$6$ & $1.07 \times 10^{3}$ & $1.40 \times 10^{-4}$ & $0.10$ & $4.63$ & $1.68$\\
			$7$ & $1.79 \times 10^{3}$ & $1.57 \times 10^{-4}$ & $0.11$ & $4.57$ & $1.63$\\
			$8$ & $2.79 \times 10^{3}$ & $1.73 \times 10^{-4}$ & $0.11$ & $4.12$ & $1.59$\\
			$9$ & $4.09 \times 10^{3}$ & $1.88 \times 10^{-4}$ & $0.12$ & $4.22$ & $1.54$\\
			$10$ & $5.86 \times 10^{3}$ & $2.07 \times 10^{-4}$ & $0.12$ & $4.19$ & $1.50$\\
			$12$ & $1.03 \times 10^{4}$ & $2.32 \times 10^{-4}$ & $0.13$ & $4.29$ & $1.43$\\
			$15$ & $1.98 \times 10^{4}$ & $2.59 \times 10^{-4}$ & $0.14$ & $4.43$ & $1.31$\\
			$20$ & $4.46 \times 10^{4}$ & $3.00 \times 10^{-4}$ & $0.15$ & $4.35$ & $1.15$\\
			$25$ & $7.97 \times 10^{4}$ & $3.28 \times 10^{-4}$ & $0.15$ & $4.32$ & $0.99$\\
			$30$ & $1.23 \times 10^{5}$ & $3.46 \times 10^{-4}$ & $0.15$ & $4.36$ & $0.93$\\
			\bottomrule
    		\end{tabular}
	\vspace*{-0.5cm}
	\end{minipage}
\end{table}

\begin{figure}
\centering
\includegraphics[width=\columnwidth]{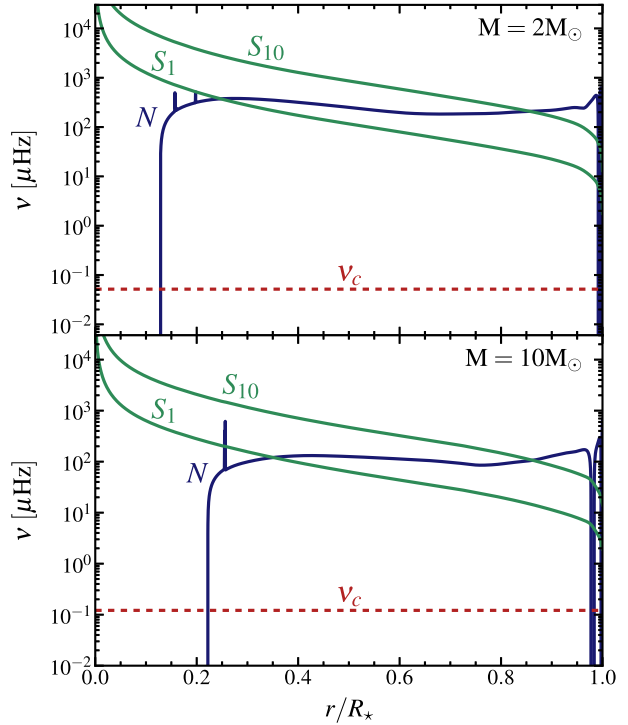}
\caption[Propagation diagrams for ZAMS models]{Propagation diagrams for 2 \Msun{} (upper panel) and 10 \Msun{} (lower panel) ZAMS models. The solid blue line shows the \brunt frequency ($N$), the solid green lines the $\ell = 1, 10$ Lamb frequency ($S_{\ell}$), and the dashed red line shows the convective turnover frequency in the core ($\nuc$).}\label{fig:pdiags}
\end{figure}

Fig. \ref{fig:pdiags} shows propagation diagrams for our 2 and 10 \Msun{} ZAMS models (upper and lower panels, respectively). Each panel shows the \brunt frequency, $N$, in the radiatively stable envelope (blue, solid line), the Lamb frequency ($S_{\ell}^{2} \equiv \ell \, (\ell + 1) \, c_{s}^{2} / r^{2}$) for $\ell = 1, 10$ (green, solid line), and the outer convective turnover frequency in the core (red, dashed line). Note the ordinate units are linear frequency, $\nu$. These diagrams highlight where g-modes and sound waves may propagate in the stellar interior, with g-modes propagating wherever $\om < N, S_{\ell}$ and sound waves where $\om > N, S_{\ell}$ \citep{aertsbook,unno}. The convective turnover frequency for all models is $\sim 3$ orders-of-magnitude less than the \brunt frequency in the radiative zone. Thus, modes with $\om \gtrsim \omc$ propagate as high-order (short-wavelength) g-modes in the radiative zone. Lastly, the contrast between panels shows the growth of the convective core in fractional radius (and mass) with increasing stellar mass and the structural similarity among main sequence models with core convection zones. 


\section{Gravity Wave Physics} \label{sec:waves}

To identify potentially observable g-modes in massive stars, we must first determine whether waves excited at the convective-radiative boundary reach the stellar surface. Waves excited at the boundary travel through the radiative zone, where they are subject to damping by radiative diffusion. The radiative damping rate for a traveling g-mode is given by
\begin{align} \label{eqn:gammarad}
	&\grad (\om, \ell, r) = K_{\rm rad} (r) \, k_{r}^{2},\\ \label{eqn:Krad}
	&K_{\rm rad} (r) = \frac{16 \, \sigma \, T(r)^{3}}{3 \, \rho(r)^{2} \kappa(r) \, c_{p}(r)}, \\ \label{eqn:kr}
	&k_{r} \approx \frac{\Lambda}{r} \frac{N}{\om}.
\end{align}
where $\Lambda^{2} \equiv \ell \, (\ell + 1)$ and $k_{r}$ is the radial wavenumber. The last approximation (eqn. \ref{eqn:kr}) relies on the WKB dispersion relation for g-modes, which have $\om \ll N, S_{\ell}$. 

We write a wave ``optical depth'' \citep[as in][]{kumar1997}
\be \label{eqn:tau}
	\tauw(\om, \ell, r) = \int_{\rm r_{conv}}^{r} dr^{\prime} \, \frac{\gamma_{\rm rad}(\om, \ell, r^{\prime})}{\vg (\omega, \ell, r^{\prime})}, 
\ee
where
\be 
	\vg \approx \frac{\omega}{k_{r}} \approx \frac{\omega^{2}\,r^{\prime}}{\Lambda \, N}.
\ee

Waves of a given frequency and degree deposit most of their energy at the radius where $\tauw \sim 1$. As $\grad$ is inversely proportional to frequency, low frequency waves damp well inside the star while high frequency waves propagate to the surface. The \textit{minimum} frequency for standing g-modes satisfies the condition $\tauw (\om, \ell, r_{outer}) \lesssim 1$, where $r_{outer}$ is the radius of the wave's outer turning point where $\om = \min \left( N, S_{\ell}\right)$. 

\begin{figure}
\centering
\includegraphics[width=\columnwidth]{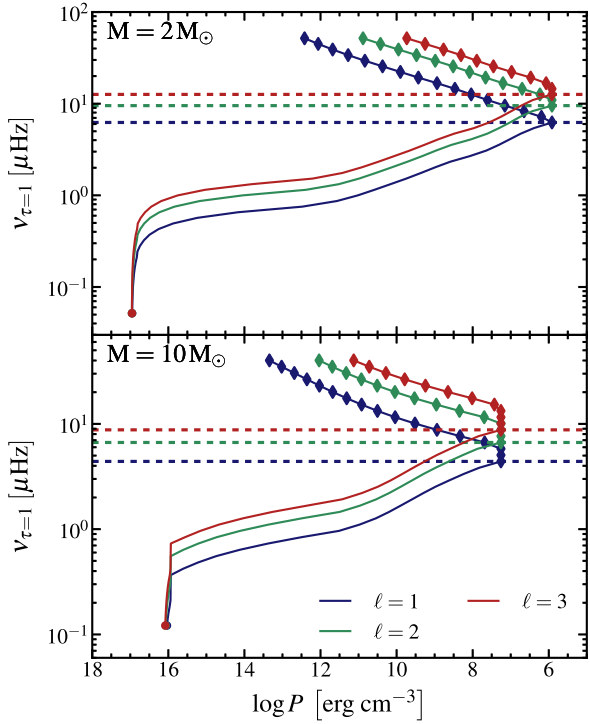}
\caption[Damping Depth for g-modes]{Location, in log pressure units, where an g-mode of frequency $\nu$ reaches unity optical depth due to radiative damping (see eqn. \ref{eqn:tau}) or its outer turning point, for 2 \Msun{} (upper panel) and 10 \Msun{} (lower panel) ZAMS models. Filled circles show the characteristic convective frequency in the convective core, diamonds mark frequencies with $\tauw < 1$ at the outer turning point (where they reflect and can set up standing waves), and the horizontal dashed lines give the minimum standing wave frequency for each $\ell$. Note that the outer turning point moves inward at high frequency as it is defined by where $\om = S_{\ell}$. Colors in each panel correspond to $\ell$, as described in the legend in the lower panel.}\label{fig:rdamp}
\end{figure}

Fig. \ref{fig:rdamp} shows the location where a wave of frequency $\nu$ has an optical depth, $\tauw \sim 1$, for g-modes with $\nu \geq \nuc$ in our 2 and 10 \Msun{} models. The horizontal dashed lines show the minimum frequency for standing g-modes, separating the low frequency waves that damp in the radiative envelope from those that reach their outer turning point and set up standing oscillations. Also shown are the characteristic turnover frequency in the convective core ($\nuc$, filled circles), and the outer turning points of g-modes with $\tauw (\om, \ell, r_{outer}) < 1$ (filled diamonds). At high frequencies, the outer turning point moves inward with increasing frequency since it is defined by where $\om = S_{\ell}$ (see Fig. \ref{fig:pdiags}). 




\section{Quasi-Adiabatic Stellar Oscillations} \label{sec:quasiad}

For each model, we use the publicly available ADIPLS\footnote{Distributed with MESA and separately at http://users-phys.au.dk/jcd/adipack.n/} adiabatic stellar oscillation package to determine the eigenfunctions for g-modes above the minimum frequency for standing waves. We then calculate the flux perturbation and growth/damping rates, in the quasi-adiabatic limit, according to the procedure described in \citet[][eqns. 1 -- 10]{shiode2012}. 

Our quasi-adiabatic calculations are only valid in the regime where the thermal time of the overlying layers is greater than the mode period. Thus, we approximate surface perturbations by their value in the vicinity of the adiabatic cutoff, $R_{ad}$, where the thermal time in the overlying layer is equal to the mode period.\footnote{In detail, we take the median value of the perturbation in the range where this ratio is between 0.1 and 10. In our 10 \Msun{} ZAMS model, the cutoff occurs at $\log P = 8.64, 8.46, 8.18, 7.73 \, \erg \cm^{-3}$ for our $n = -16, -10, -5, -1$ g-modes at frequencies of $\nu = 4.38, 6.73, 12.79, 39.85 \muHz$.}  Comparison with non-adiabatic growth/damping rates and eigenfunctions (Townsend \& Teitler \textit{in prep}) shows that our quasi-adiabatic calculations provide a good approximation for damping rates and surface amplitudes to within factors of a few.

The short wavelength g-modes predominantly excited by convection are subject to large radiative damping. Those able to set up global standing waves are still linearly damped with lifetimes of $\sim 1 - 10^{5} \, \yr$ (with the exception of some linearly driven modes). Damping rates are fit reasonably well by power laws in mode frequency, with indices of $-3$ to $-8$, where flatter power laws apply for more evolved models. Deviations from the power-law arise primarily from opacity effects in the stellar envelope, whose relative importance depends on mode frequency. 

The convective excitation rate and mode lifetime together set the equilibrium mode energy 
\be
	\Emode (\om, \ell) = \frac{1}{2} \, \frac{d \Edotg}{d \ln \om \, d \ell} \Nm (\om, \ell)^{-1} \gamma^{-1}, \label{eqn:Emode}
\ee
where $\Emode \equiv \omega^2/2 \, \int_0^{M_{ad}} \left[\xi_r^2 + \Lambda^{2}\xi_h^2 \right] dm$, $\Nm (\om, \ell)$ is the number of modes in a logarithmic bin in $\om$ (at fixed $\ell$), and $\gamma$ is the damping rate of the mode as determined by all non-adiabatic effects (radiative diffusion, nuclear driving, and any convective viscosity). For g-modes with $\om \ll N$, $\Nm (\om, \ell) \approx (n + \ell/2) (2\,\ell + 1)$, where $n$ is the number of radial nodes in the mode eigenfunction \citep{unno}. 

In linear adiabatic stellar oscillation theory, the amplitude of the temperature and density perturbations, and the magnitude of the fluid displacement, are arbitrary.  We convert the linear theory results into realistic predictions by normalizing the g-mode eigenfunctions using the estimate of the mode energy in equation \ref{eqn:Emode}.  We then use these properly normalized eigenfunctions to calculate the disk integrated perturbations as follows.


\subsection{Surface perturbations} \label{ssec:surf}

For modes with $\ell \geq 1$, disk-averaging effects reduce the observable amplitude of oscillations as neighboring surface elements oscillate out of phase with one another. For each mode, we calculate the disk-integrated, limb-darkened, quasi-adiabatic perturbation to the bolometric magnitude and surface radial velocity for each stellar model \citep[as in][]{dziembowski1977}:
\begin{align}
	\left| \delta m \right| = & 1.087 \left[ b_{\ell} \, \frac{\delta F}{F_{0}} + 
		\left(2 b_{\ell} - c_{\ell}\right) \frac{\xi_{r}}{R}\right]_{r \approx R_{ad}},\\
	\left| \delta \vdisk \right| = & \left[u_{\ell} \, \xi_{r} \, \omega + v_{\ell} \, \xi_{h} \, \omega \right]_{r \approx R_{ad}},
\end{align}
where $b_{\ell}, c_{\ell}, u_{\ell},$ and $v_{\ell}$ are coefficients related to the effect of limb-darkening on the visibility of modes of a given $\ell$, $\xi_{r,h}$ are the radial and horizontal displacement eigenfunctions, and $\delta F / F_{0}$ is the fractional perturbation to the radiative flux due to the oscillation. We assume Eddington limb-darkening for simplicity, and thus use the $b_{\ell}, c_{\ell}, u_{\ell},$ and $v_{\ell}$ coefficients tabulated in \citet{dziembowski1977}. 

We primarily use $\delta \vdisk$ to compare our calculations with previous estimates and observational constraints on solar g-modes (see \S \ref{ssec:solar}). For the more massive stellar models, which we show produce a much smaller velocity signal, we also consider a simpler estimate of the non-disk-averaged total velocity perturbation
\be 
	\left| \delta v_{\rm tot} \right| = \frac{1}{2} \, \left[\left|\xi_{r} \omega \right|^{2} + \left|\xi_{h} \omega\right|^{2}\right]_{r \approx R_{ad}}^{1/2},
\ee
which might be connected to the non-thermal velocities needed to fit spectral line profiles in hot, massive stars: microturbulence and macroturbulence. 

As a general rule, the procedure described above results in a ratio of flux-to-velocity-perturbations of order $\delta m / \delta \vdisk \sim 1 \mumag/\cms$ for $\ell = 1$ modes. We find that this ratio varies between $0.1 - 10 \mumag/\cms$, with a trend towards larger values at lower frequencies. There is no clear trend with mass or stellar evolutionary state. 


\section{Results} \label{sec:res}

\subsection{Brightness perturbations} \label{ssec:dm}

We find that, unless the excitation is dominated by an overshoot region $\lesssim 10$ per cent of a pressure scale height thick, corresponding to a peak excitation frequency  of $\gtrsim 10 \, \omc$, individual oscillation modes excited by convection produce brightness fluctuations of $\delta m \lesssim 10 \, \mu{\rm mag}$ (except at the TAMS, see \S \ref{sssec:dmevol}). 

We thus focus on the effect of an ensemble of stochastically excited modes on the intrinsic variability of the star. This RMS magnitude perturbation is an incoherent sum over all modes in a given logarithmic bin in frequency:
\be \label{eqn:rmsdm}
	\rmsdm = \left[\sum_{\Delta \ln \omega, \; {\rm all} \; \ell} \; 
		\sum_{m=-\ell}^{\ell} \left[\delta m (\ell, m, \omega)\right]^{2}\right]^{1/2},
\ee
where the sum is over a logarithmic bin in $\omega$ and all $\ell$ (though practically only $\ell \lesssim 5$ contribute). 


\begin{figure*}
\centering
\includegraphics[width=\textwidth]{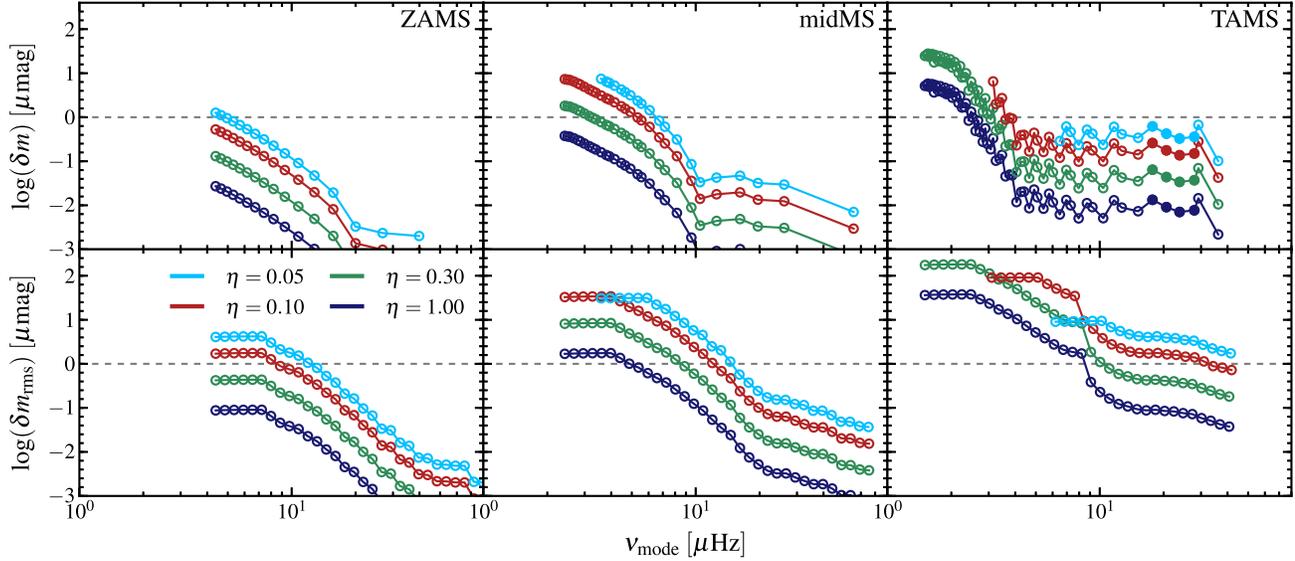}
\caption[Brightness fluctuations for 10 \Msun{} model along the main sequence]{Individual $\ell = 1$ g-mode (top panels) and RMS (lower panels; see eqn. \ref{eqn:rmsdm}) surface brightness fluctuations as a function of frequency for convectively excited g-modes in our 10 \Msun{} model at the ZAMS (left panels), midMS (middle panels), and TAMS (right panels). Individual $\ell = 2$ modes have similar amplitudes to those shown in the top panels for all phases, and $\ell = 3$ modes approach parity at the TAMS. Open circles in the top panels mark individual mode frequencies, while in the bottom panels they show frequencies at which we have calculated $\rmsdm$. Note the decrease in the minimum standing wave frequency (i.e., the low frequency cutoff of the spectrum), and corresponding increase in the magnitude of the maximum brightness perturbation, with evolution.  In our evolved models, the low frequency cutoff of the spectrum for $\eta \lesssim 0.3$ is set by the minimum excitation frequency for which our model is valid, $\nuc / \eta$, since this is larger than the minimum frequency for standing waves. There is likely to be convectively excited power at these frequencies \citep[see, e.g.,][]{rogers2006}, but it is not accounted for here.}\label{fig:dmevol}
\end{figure*}


The left panels of Fig. \ref{fig:dmevol} show the predicted $\delta m$ perturbations due to individual $\ell = 1$ modes (top) and the total $\rmsdm$ (bottom) as a function of frequency for our 10 \Msun{} ZAMS models. The open circles in the upper left panel (for $\delta m$) correspond to individual g-mode frequencies, while the circles in the bottom left panel are frequencies where $\rmsdm$ has been calculated from the individual modes. The flatter spectrum of the RMS magnitude perturbations, relative to that of the individual $\ell = 1$ g-modes, results from summing the contributions from all modes in a given logarithmic frequency bin, including those of higher $\ell$ (which are not plotted in the upper left panel; see equation \ref{eqn:rmsdm}). 

For ZAMS models, the combination of the convective excitation spectrum and that of radiative damping combine to produce a nearly flat spectrum of mode energies for the standing g-modes excited by convection, since $\dot{E} \propto \nu^{-13/2}$ and $\gamma \propto \nu^{-6}$. However, there is a strong decrease in surface amplitude with increasing frequency at fixed mode energy, which is well fit by $\delta m \propto \nu^{-4}$, due primarily to the outer turning point moving inward. This latter trend dominates the $\delta m$ spectra shown in the top panels of Fig. \ref{fig:dmevol}. The spectra of flux perturbations is qualitatively similar to that shown in Fig. \ref{fig:dmevol} for the range of stellar masses we investigated. 

\begin{figure}
\centering
\includegraphics[width=\columnwidth]{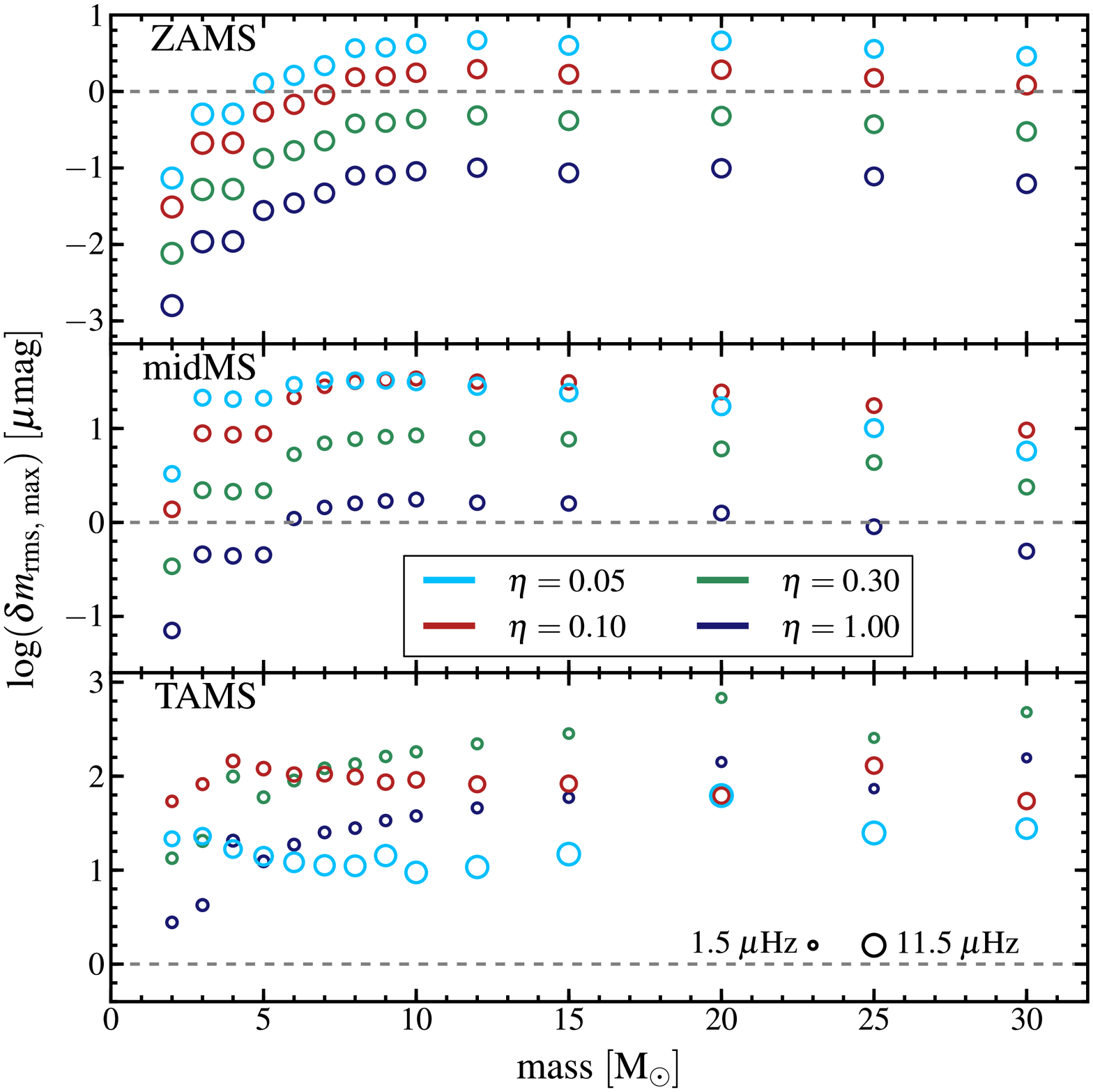}
\caption[Brightness fluctuations for all masses]{Peak RMS surface brightness fluctuations $\rmsdm$ for convectively excited g-modes, calculated according to eqn. \ref{eqn:rmsdm}, as a function of mass. Color corresponds to the value of $\eta$ and size to mode frequency, as described in legend. Panels show results for models at their labeled evolutionary states. In the evolved models, the minimum frequency for excitation at small $\eta$ ($\nuc/\eta$) is larger than the minimum frequency for standing waves ($\nu_{\rm min}$; Table 1), leading to a suppression in the total flux perturbation for the smallest values of $\eta$.}\label{fig:maxdm}
\end{figure}

The top panel of Fig. \ref{fig:maxdm} shows the maximum $\rmsdm$ in ZAMS models at each mass, with color corresponding to overshoot thickness and symbol size denoting the frequency of the peak flux perturbation. In more massive stars, the closer match between standing g-mode frequencies and the characteristic excitation frequency, and the overall larger luminosity and mach number, lead to larger brightness perturbations. However, the frequency of peak brightness, $\approx 5 \muHz$ ($\sim 0.4 \perday$), is roughly constant with increasing mass.


\subsubsection{Effect of stellar evolution} \label{sssec:dmevol}

Both the mach number in the core and the stellar luminosity increase during main sequence sequence evolution, leading to increased power input to g-modes and a larger characteristic convective frequency, $\nuc$. In addition, the minimum frequency for standing waves decreases along the main sequence as the thermal time at the mode outer turning point increases. Together these imply an increase in g-mode surface amplitudes with evolution, as well as the appearance of ever lower frequency modes. 

The top and bottom panels of Fig. \ref{fig:dmevol} show the evolution along the main sequence of the power spectrum for individual modes and $\rmsdm$, respectively, in our 10 \Msun{} model. The RMS flux perturbation (lower panels) \emph{at fixed frequency} increases with evolution, and the appearance of lower frequency modes at the surface leads to larger signals at lower frequencies than appeared on the ZAMS. 

The chemical composition gradient left behind by the receding convective core during the main sequence evolution produces an extended bump in the \brunt profile; this results in enhanced trapping behavior near the end of the main sequence as modes separate into those with most of their energy in that composition-gradient region and those primarily in the radiative zone. This effect leads to the differences in the surface amplitudes of adjacent modes seen in the upper right panel of Fig. \ref{fig:dmevol}.

As shown in Figs. \ref{fig:dmevol} and \ref{fig:maxdm}, we predict g-mode power peaked at near $5 \muHz$ on the ZAMS, but this peak approaches $\sim 1 \muHz$ at the end of the main sequence. However, in our evolved models, the minimum frequency for standing g-modes can be less than the minimum frequency for which our excitation model applies, $\omc / \eta$, when the overshoot region is $\lesssim 30$ per cent of a pressure scale height. This is shown clearly in the low-frequency cutoff of the spectra in Fig. \ref{fig:dmevol}, and in the lower panels of Fig. \ref{fig:maxdm}, where the magnitude of the maximum perturbation for $\eta = 0.05, 0.1$ is smaller than that for larger $\eta$ and appears at a higher frequency. There is likely convective excitation power at these low frequencies, but it is not accounted for in our model \citep[see, e.g.,][]{rogers2006}. 

If we assume the energy flux into modes below $\nuc/\eta$ is constant, as appears to be the case in the \citet{rogers2006} simulations, the equilibrium energy in modes below this cutoff scales as $\sim \nu^{4}$. However, a strong scaling of surface amplitude at fixed mode energy, as we find in the case of our 10 \Msun{} TAMS model, can lead to an increase in the maximum $\rmsdm$ by a factor of a few and a shift of the peak to slightly lower frequencies than appear in the lower panels of Figs. \ref{fig:dmevol} and \ref{fig:maxdm}. At yet lower frequencies, however, approaching the minimum for standing waves, $\nu_{min}$, the predicted flux perturbations decrease. 


For \textit{Kepler} and CoRoT observations with micromagnitude precision photometry, \emph{the RMS brightness fluctuations contributed by convectively excited g-modes should be detectable during the main sequence for all stars considered}, if the driving is dominated by a convective overshoot region with width $\lesssim 30$ per cent of a pressure scale height at the top of the core. The expected amplitudes reach $\sim 100{\rm s \; of \;} \mu$mag around $\nu \sim 1 - 10 \muHz$ ($0.08 - 0.8 \perday$). 

\subsection{Velocity perturbations} \label{ssec:dv}

In addition to brightness perturbations, non-radial modes produce velocity perturbations which may be observed either in the disk-integrated radial velocity signature or via the motions they generate where spectral lines are formed \citep[``microturbulence''; e.g.,][and references therein]{cantiello2009}. In massive main sequence stars, we are interested in the latter case; moreover, we are concerned with the aggregate effect of many modes, as discussed above for the brightness perturbations.

\begin{figure}
\centering
\includegraphics[width=\columnwidth]{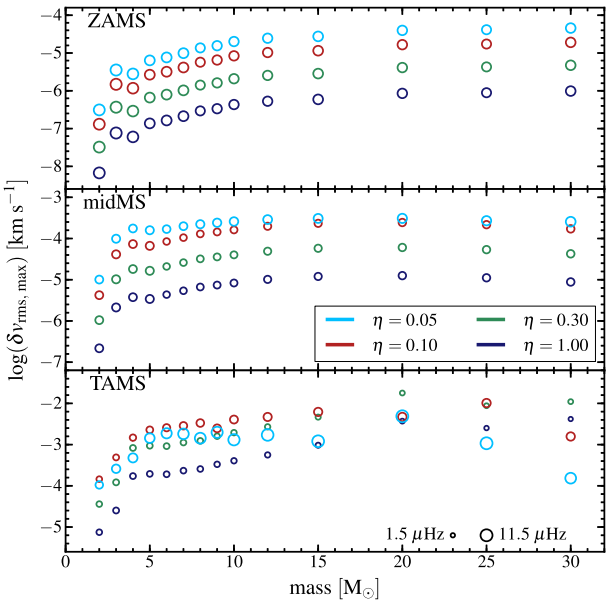}
\caption[Brightness fluctuations for all masses]{Peak RMS surface velocity fluctuations convectively excited g-modes as a function of mass. Symbols and colors are as in Fig. \ref{fig:maxdm}.}\label{fig:maxdv}
\end{figure}

In Fig. \ref{fig:maxdv}, we show the maximum predicted RMS velocity fluctuations as a function of initial stellar mass and evolutionary phase. For g-modes excited by the convective core, we predict surface velocity fluctuations that are always $\ll 1 \kms$, whereas the observed non-thermal surface velocities in massive main sequence stars are $\gtrsim 1 - 10 \kms$. Even in our most optimistic scenario for excitation, with $\eta = 0.05$, g-modes excited by core convection do not produce sufficiently large velocities to explain the observed microturbulence.


\subsection{Observability of solar g-modes} \label{ssec:solar}

We have also computed a solar model according to modern asteroseismology constraints, in order to determine the surface velocities of solar g-modes implied by our excitation model. If the excitation is dominated by the highly-stratified solar convection zone, g-modes of a given frequency $\omega$ are primarily excited at the place in the convection zone where  $\om \sim \omc(r)$ \citep{kumar1999}.   Thus  high frequency g-modes are predominantly excited near the solar surface, while low frequency g-modes are excited near the radiative-convective boundary in the interior.   Assuming the convection zone behaves like a polytrope with index $n \approx 1.5$, equation \ref{eqn:spect} still applies, but with  $\lmax = r / H \, (\om / \omc)^{2/3}$, where $r$ and $H$ are evaluated at the interior radiative-convective boundary where g-modes begin to propagate, $a = 7/2$, $b = 2$, and $\eta = 1$.   

\begin{figure}
\centering
\includegraphics[width=\columnwidth]{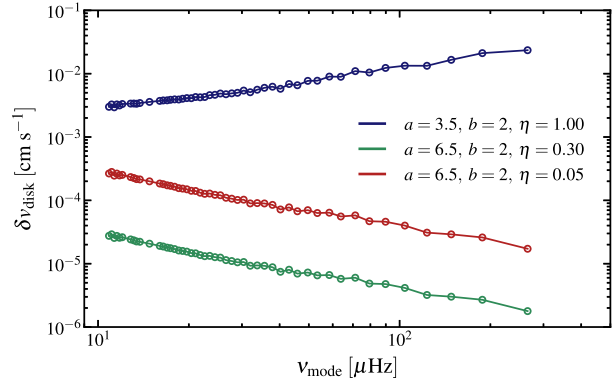}
\caption[Disk-integrated velocity perturbations for solar g-modes]{Disk-integrated velocity perturbations for $\ell = 1$ solar g-modes. For comparison, $\ell = 2, 3$ modes have smaller velocity perturbations by factors of $\sim 3$ and 10, respectively. The blue points show the amplitudes of modes assuming that the highly-stratified solar convection zone dominates the exciation, as compared to the green and red points which give the result of excitation by an overshoot region of 30 or 5 per cent of a pressure scale height, respectively.}\label{fig:solarv}
\end{figure}

However, if the excitation is instead predominantly due to a thin overshoot region of width $\eta H$ at the base of the convection zone, then we assume the excitation spectrum at fixed position ($a = 13/2$), with $\eta < 1$. In Fig. \ref{fig:solarv}, we show the disk-integrated g-mode velocity amplitudes that result from using the convection-zone-dominated and overshoot-dominated excitation spectra. In the overshoot case, we choose values $\eta = 0.05, 0.3$ that span the range suggested by helioseismic investigations and numerical simulations of solar convection \citep[e.g.,][]{dalsgaard2011,rogers2005,rogers2006}.

In all reasonable excitation scenarios, we find amplitudes of $\delta \vdisk \lesssim 0.3 \mms$, which are consistent with those derived by \citet{kumar1996}. Our results are also consistent with the observational upper limits that place $\delta \vdisk \lesssim 10 \mms$ at $\sim 100 \muHz$ \citep{appourchaux2010}. 

The intensity-to-velocity ratio we find for these high frequency solar g-modes is $\delta m / \delta v \lesssim 1 \mumag / \cms$, so that typical intensity perturbations are $\lesssim 10^{-2} \mumag$. This is also well below the published upper limits from \citet{appourchaux2000} of $\sim 0.5 \mumag$.


\section{Discussion} \label{sec:disc}

Using a simple parametrization for the convective excitation of gravity modes, we predict that the convective cores of stars with masses $\gtrsim 2$ \Msun{} will excite observable surface brightness fluctuations while on the main sequence. If the excitation is dominated by an overshoot region having a width of $\lesssim 30$ per cent of the pressure scale height at the top of the convective core, these stars will exhibit intrinsic photometric variability with amplitudes up to 10s of micromagnitudes at frequencies of $5 - 10 \, \muHz$ ($0.4 - 0.8 \perday$) on, or just red of the solar metallicity ZAMS.  These flux variations are largest for massive stars with $M \gtrsim 5$ \Msun{} (Fig. \ref{fig:maxdm}).  

As the stars evolve along the main sequence, we predict a strong increase in the flux variability, reaching 100s of micromagnitudes at frequencies $\lesssim 10 \muHz$, along with the appearance of power at ever lower frequencies, down to $\sim 1 \muHz$ ($0.08 \perday$) at the TAMS. We predict no observable accompanying velocity signal. 

Given the relatively low amplitudes of individual g-modes, we have framed our predictions in terms of the RMS flux variations at a given frequency, rather than the amplitudes of individual stellar normal modes.  The latter are smaller by factors of $\sim 3-30$ at these frequencies (with larger corrections at lower frequencies), making it challenging to detect individual normal modes. However, as the duration of observation for \textit{Kepler} and CoRoT stars increases, it may become possible to detect individual normal modes in a periodogram analysis, since the long lifetimes of these modes ensures that they will become more prominent in longer duration observations \citep[e.g.][]{dupret2009}.

Our excitation model agrees with previous predictions that the surface amplitudes of solar g-modes are $\lesssim 0.3 \mms$ (Fig. \ref{fig:solarv}), thus implying that \emph{photometry of massive main sequence stars provides the best potential window into the convective excitation of g-modes.}

\subsection{Theoretical uncertainties} \label{ssec:TheoryUncert}

\subsubsection{Overshoot excitation} \label{ssec:over}

Detailed asteroseismic modeling provides a best fit width for the solar overshoot region (``tachocline'') of between 5 and 40 per cent of a pressure scale height at the base of the convection zone, depending on the overshoot prescription used \citep[][and references therein]{dalsgaard2011}.  Numerical simulations by \citet{rogers2005, rogers2006} imply an overshoot thickness at the small end of this range $\sim 0.05 H$. However, asteroseismic results from other solar-like stars suggest that the thickness of the overshoot layer  may not be universal even among similar stellar types \citep[e.g.,][]{lebreton2012}.

For main sequence stars with core convection zones, numerical simulations by \citet*{browning2004} find that the overshoot region has, in our terms, $\eta \lesssim 0.2$. Results from asteroseismic modeling of massive main sequence pulsators are broadly consistent with overshoot of a few tenths of a pressure scale height, but show a range of best fit parameters from consistent with zero \citep[e.g.,][]{aerts2011} to more than 40 per cent of a pressure scale height \citep[e.g.,][]{briquet2007}; further corroboration is found in isochrone fitting of open clusters \citep[][and references therein]{brott2011}. This broad base of observational results, in combination with arguments presented here, suggest that g-mode excitation by convective overshoot could produce observable surface brightness perturbations in massive main sequence stars. 


\subsubsection{Effects of rotation on g-modes} \label{sssec:rot}

Using typical observed surface velocities for massive stars \citep[e.g.,][]{wolff2006}, the corresponding rotation frequencies are given by
\be 
\Omega \approx 3.5 \muHz \; \left(\frac{v}{150 \kms}\right) \left(\frac{R}{\rsun}\right)^{-1}, 
\ee
i.e., $\Omega \simeq 3 \, (1) \, \muHz$ for $M = 2 \, (10) M_\odot$.

The effect of the Coriolis force on g-mode propagation can be seen in the WKB dispersion relation \citep[as in e.g.,][eqn. 21]{kumar1999}:
\be
\omega^{2} \approx 4 \frac{\left({\bf k \cdot \Omega}\right)^2}{k^2} + N^{2} \frac{\left({\bf k \cdot {\hat g}}\right)^2}{k^2},
\ee
Thus, g-modes only propagate where $\omega > 2\, (\Omega \cdot \hat{k})$, and modes with frequencies $\omega \lesssim \Omega$ are confined to the equator.   

As discussed in \S \ref{sec:res}, much of the convectively excited power is in modes with frequencies of $\sim 1-5 \muHz$, comparable to the median rotation frequencies of massive main sequence stars.  Thus we expect that rotation will have a non-negligible effect on the surface manifestation of convectively excited g-modes (it may also change the excitation of g-modes since rapidly rotating convection has different statistical properties than non-rotating convection).    This is particularly true as the star evolves to the TAMS.    Given the potential importance of rotation, it would be particularly interesting to compare high precision photometry of slowly vs. rapidly rotating main sequence A-O stars.

Additionally, if the surface of the star is rotating more rapidly than the core, the star may possess critical layers that significantly increase the radiative damping of outgoing g-modes and preclude the establishment of standing waves. Such critical layers may in fact be set up by the angular momentum carried by the g-modes themselves \citep*[see e.g.,][]{rogers2012}. 

\subsubsection{Effects of stellar model parameters} \label{ssec:StEvo}

To assess the effects of stellar  modeling parameters on our predictions, we evolved a 10 \Msun{} star with variable mixing parameters and rotation. We tested mixing length parameters $\alpha_{\rm MLT} = 1, 1.5, 2$; boundaries determined by the Schwarzschild and Ledoux criteria without overshoot; and the Schwarzschild criterion with 30 per cent overshoot and initial surface rotation velocities of $50, 100, 200 \kms$ (with and without the chemical mixing induced by Taylor-Spruit magnetic fields).

We find that the uncertainty in our predicted mode amplitudes at fixed central hydrogen mass fraction due to stellar evolutionary parameters may be up to a factor of $\sim 5$, with the models presented in our figures roughly corresponding to the median of the distribution. Changing $\alpha_{\rm MLT}$ produces the largest effect, as larger $\alpha_{\rm MLT}$ corresponds to smaller $\omc$ and larger $\Mc$; the  overall normalization of the convective excitation spectrum scales as $\alpha_{\rm MLT}^{-2}$ for $a = 13/2$ (see eqn. \ref{eqn:spect}). Convective boundary definitions can also change the ratio of $\om_{\rm min} / \omc$ by up to a few percent on the ZAMS, changing the fraction of the convectively excited g-mode power that reaches the stellar surface by $\lesssim 25$ per cent. By the end of the main-sequence these effects can amount to an order of magnitude dispersion among models, as the determination of convective boundaries affects the progression of stellar evolution.  

\subsubsection{Turbulent power spectrum} \label{sssec:turbspec}

In their study of the detectability of solar g-modes, \citet{belkacem2009} employ a Lorentzian eddy-time-correlation function for coupling the turbulent convective motions to g-mode excitation. They find that this results in a 30-fold increase in solar g-mode amplitudes, as compared to the assumptions used in our analysis (and previous work; e.g., \citealt{kumar1996}). \citet{samadi2010} also employed this formalism to study convectively excited g-modes in 10, 15, and 20 \Msun{} main sequence models having $X_{c} = 0.5$. They find amplitudes of $\sim 10 \mumag$ for individual $\ell = 1$ g-modes in these stars, also $\sim$ 30 times larger than the amplitudes in our $\eta = 1$ case (see our Fig. \ref{fig:dmevol} and their fig. 2).  


\subsubsection{Near-surface convection zones} \label{ssec:surfcz}

 Massive main sequence stars of near solar metallicity have vigorous near-surface convection zones due to the iron opacity bump \citep{cantiello2009}.  These convection zones are another potentially important source of g-modes near the stellar surface, having mach numbers of $\gtrsim 0.01$. Near-surface convection zones have characteristic frequencies of 10s to 100s of $\muHz$, but the small scale-heights near the stellar surface imply that nearly all of the convective power is at larger characteristic $\ell \gtrsim 30$. These high $\ell$ g-modes are unlikely to produce any significant surface brightness perturbations, but, as suggested by \citet{cantiello2009}, the velocity field associated with these waves might account for the micro/macro-turbulent velocity fields inferred via spectroscopic modeling.   We leave a detailed investigation of the excitation and propagation of these modes to future work.

\subsection{Observational prospects} \label{ssec:obs}

\subsubsection{Distinguishing stochastically excited modes} \label{sssec:disting}

Some of the stars we have investigated lie in the SPB instability strip, where stars are observed to pulsate in low-degree g-modes with frequencies of $0.6 - 3 \perday$.  These linearly excited modes have frequencies very near the peak of the stochastically excited modes.  It is thus important to determine how to observationally distinguish between linearly and stochastically excited modes. 

Unlike p-modes stochastically excited by envelope convection zones (i.e., solar-like oscillations), the convectively excited modes explored in this work have linear damping timescales (primarily due to radiative diffusion) of years or longer, much longer than a typical observation timescale. Thus the spectral broadening and time-frequency diagram characteristics that distinguish stochastically-driven modes in the former case cannot be leveraged here \citep[e.g.,][]{bedding2011,belkacem2009science}. We must instead rely on the fact that convectively excited g-modes should be observable in stars outside the linear instability strips and at frequencies that should otherwise be damped in linear analyses.

\subsubsection{Have convectively excited g-modes already been detected?} \label{sssec:lit}

\citet{aerts2009} have shown that the ``macroturbulent'' velocity fields observed in spectra of blue supergiants can be explained by the collective effect of an ensemble of excited g-modes, which span a similar frequency range to that presented here. More recent ongoing work by \citet{simon-diaz2011}, has shown that macroturbulence is present in O and B dwarf stars as well, suggesting that perhaps the same low-frequency pulsations are present on the main sequence. We have shown however, that if individual mode amplitudes are set by the competition between convective excitation by the core and radiative damping in the envelope (see eqn. \ref{eqn:Emode}), the aggregate effect is too small to explain the observed macroturbulence (see Fig. \ref{fig:maxdv}). The source of these turbulent motions could instead be g-modes excited by near-surface convection zones, as discussed in \S \ref{ssec:surfcz} above and \citet{cantiello2009}. Another interesting possible source is the photon bubble instabilities that occur in the atmospheres of massive stars \citep{turner2004}.

There is tantalizing evidence for convectively excited g-modes in recent observations of A, B and O stars with {\it Kepler} \citep{balona2011,uytterhoeven2011,blomme2011}. In their study of B-star pulsators, \citet{balona2011} find evidence for theoretically unexpected, low frequency pulsations in SPB stars and SPB/$\beta$ Cep hybrid pulsators. Some of these stars lie outside the classical instability strips for low-order modes, potentially implying that a previously unexplored driving mechanism is at work in these stars.

In studies of O-stars, \citet{blomme2011} have found an unexpected ``red-noise'' component to the stellar photometric power spectrum, even after instrumental corrections have been applied. This noise appears at the frequencies and amplitudes that agree with our predictions; however, the inferred mode lifetimes of hours to days for the ``red-noise'' modes are incommensurate with those of the g-modes explored here, which have lifetimes of years to Myrs.  

As a general comment, we note that many studies remove or disregard power at low-frequencies below $0.5$ or $0.2 \perday$, depending on the study. They appeal to instrumental effects both known and un-characterised. However, we urge caution:   the complete removal of this low frequency power may take with it the signatures of convectively excited g-modes.

\section*{Acknowledgments}

We acknowledge a stimulating workshop at Sky House where these ideas germinated.  We would like to thank Bill Paxton for invaluable support with MESA, Rich Townsend for providing non-adiabatic oscillation modes for comparison with our quasi-adiabatic calculations, and Conny Aerts and Neal Turner for helpful feedback that improved this work. JS would also like to thank the American Museum of Natural History for providing office space and support during the completion of this work. This work was partially supported by a Simons Investigator award from the Simons Foundation to EQ, the David and Lucile Packard Foundation, and the Thomas and Alison Schneider Chair in Physics at UC Berkeley. Additional support was provided by the National Science Foundation under grants PHY 11-25915 and AST 11-09174, and by NASA Headquarters under the Earth and Space Science Fellowship Program - Grant 10-Astro10F-0030.

\vspace*{-0.1cm}

\footnotesize{
	\bibliographystyle{mn2e}
	\bibliography{jhsrefs}

\begin{thebibliography}{49}
\expandafter\ifx\csname natexlab\endcsname\relax\def\natexlab#1{#1}\fi

\bibitem[{{Aerts} {et~al}\mbox{.}(2011){Aerts}, {Briquet}, {Degroote}, {Thoul},
  \& {van Hoolst}}]{aerts2011}
{Aerts} C., {Briquet} M., {Degroote} P., {Thoul} A., {van Hoolst} T., 2011,
  \aap, 534, A98

\bibitem[{{Aerts} {et~al}\mbox{.}(2010){Aerts}, {Christensen-Dalsgaard}, \&
  {Kurtz}}]{aertsbook}
{Aerts} C., {Christensen-Dalsgaard} J., {Kurtz} D.~W., 2010,
  {Asteroseismology}, 1st edn. Springer, p. 866

\bibitem[{{Aerts} {et~al}\mbox{.}(2009){Aerts}, {Puls}, {Godart}, \&
  {Dupret}}]{aerts2009}
{Aerts} C., {Puls} J., {Godart} M., {Dupret} M.-A., 2009, \aap, 508, 409

\bibitem[{{Appourchaux} {et~al}\mbox{.}(2010){Appourchaux}, {Belkacem},
  {Broomhall}, {Chaplin}, {Gough}, {Houdek}, {Provost}, {Baudin}, {Boumier},
  {Elsworth}, {Garc{\'{\i}}a}, {Andersen}, {Finsterle}, {Fr{\"o}hlich},
  {Gabriel}, {Grec}, {Jim{\'e}nez}, {Kosovichev}, {Sekii}, {Toutain}, \&
  {Turck-Chi{\`e}ze}}]{appourchaux2010}
{Appourchaux} T. {et~al.}, 2010, \aapr, 18, 197

\bibitem[{{Appourchaux} {et~al}\mbox{.}(2000){Appourchaux}, {Fr{\"o}hlich},
  {Andersen}, {Berthomieu}, {Chaplin}, {Elsworth}, {Finsterle}, {Gough},
  {Hoeksema}, {Isaak}, {Kosovichev}, {Provost}, {Scherrer}, {Sekii}, \&
  {Toutain}}]{appourchaux2000}
{Appourchaux} T. {et~al.}, 2000, \apj, 538, 401

\bibitem[{{Auvergne} {et~al}\mbox{.}(2009){Auvergne}, {Bodin}, {Boisnard},
  {Buey}, {Chaintreuil}, {Epstein}, {Jouret}, {Lam-Trong}, {Levacher},
  {Magnan}, {Perez}, {Plasson}, {Plesseria}, {Peter}, {Steller}, {Tiph{\`e}ne},
  {Baglin}, {Agogu{\'e}}, {Appourchaux}, {Barbet}, {Beaufort}, {Bellenger},
  {Berlin}, {Bernardi}, {Blouin}, {Boumier}, {Bonneau}, {Briet}, {Butler},
  {Cautain}, {Chiavassa}, {Costes}, {Cuvilho}, {Cunha-Parro}, {de Oliveira
  Fialho}, {Decaudin}, {Defise}, {Djalal}, {Docclo}, {Drummond}, {Dupuis},
  {Exil}, {Faur{\'e}}, {Gaboriaud}, {Gamet}, {Gavalda}, {Grolleau}, {Gueguen},
  {Guivarc'h}, {Guterman}, {Hasiba}, {Huntzinger}, {Hustaix}, {Imbert},
  {Jeanville}, {Johlander}, {Jorda}, {Journoud}, {Karioty}, {Kerjean},
  {Lafond}, {Lapeyrere}, {Landiech}, {Larqu{\'e}}, {Laudet}, {Le Merrer},
  {Leporati}, {Leruyet}, {Levieuge}, {Llebaria}, {Martin}, {Mazy}, {Mesnager},
  {Michel}, {Moalic}, {Monjoin}, {Naudet}, {Neukirchner}, {Nguyen-Kim},
  {Ollivier}, {Orcesi}, {Ottacher}, {Oulali}, {Parisot}, {Perruchot},
  {Piacentino}, {Pinheiro da Silva}, {Platzer}, {Pontet}, {Pradines},
  {Quentin}, {Rohbeck}, {Rolland}, {Rollenhagen}, {Romagnan}, {Russ}, {Samadi},
  {Schmidt}, {Schwartz}, {Sebbag}, {Smit}, {Sunter}, {Tello}, {Toulouse},
  {Ulmer}, {Vandermarcq}, {Vergnault}, {Wallner}, {Waultier}, \&
  {Zanatta}}]{corot}
{Auvergne} M. {et~al.}, 2009, \aap, 506, 411

\bibitem[{{Balona} {et~al}\mbox{.}(2011){Balona}, {Pigulski}, {Cat}, {Handler},
  {Guti{\'e}rrez-Soto}, {Engelbrecht}, {Frescura}, {Briquet}, {Cuypers},
  {Daszy{\'n}ska-Daszkiewicz}, {Degroote}, {Dukes}, {Garcia}, {Green}, {Heber},
  {Kawaler}, {Lehmann}, {Leroy}, {Molenda-{\.Z}aaowicz}, {Neiner}, {Noels},
  {Nuspl}, {{\O}stensen}, {Pricopi}, {Roxburgh}, {Salmon}, {Smith},
  {Su{\'a}rez}, {Suran}, {Szab{\'o}}, {Uytterhoeven}, {Christensen-Dalsgaard},
  {Kjeldsen}, {Caldwell}, {Girouard}, \& {Sanderfer}}]{balona2011}
{Balona} L.~A. {et~al.}, 2011, \mnras, 413, 2403

\bibitem[{{Beck} {et~al}\mbox{.}(2011){Beck}, {Bedding}, {Mosser}, {Stello},
  {Garcia}, {Kallinger}, {Hekker}, {Elsworth}, {Frandsen}, {Carrier}, {De
  Ridder}, {Aerts}, {White}, {Huber}, {Dupret}, {Montalb{\'a}n}, {Miglio},
  {Noels}, {Chaplin}, {Kjeldsen}, {Christensen-Dalsgaard}, {Gilliland},
  {Brown}, {Kawaler}, {Mathur}, \& {Jenkins}}]{beck2011sci}
{Beck} P.~G. {et~al.}, 2011, Science, 332, 205

\bibitem[{{Bedding}(2011)}]{bedding2011}
{Bedding} T.~R., 2011, ArXiv e-prints

\bibitem[{{Bedding} {et~al}\mbox{.}(2011){Bedding}, {Mosser}, {Huber},
  {Montalb{\'a}n}, {Beck}, {Christensen-Dalsgaard}, {Elsworth},
  {Garc{\'{\i}}a}, {Miglio}, {Stello}, {White}, {De Ridder}, {Hekker}, {Aerts},
  {Barban}, {Belkacem}, {Broomhall}, {Brown}, {Buzasi}, {Carrier}, {Chaplin},
  {di Mauro}, {Dupret}, {Frandsen}, {Gilliland}, {Goupil}, {Jenkins},
  {Kallinger}, {Kawaler}, {Kjeldsen}, {Mathur}, {Noels}, {Aguirre}, \&
  {Ventura}}]{bedding2011a}
{Bedding} T.~R. {et~al.}, 2011, \nat, 471, 608

\bibitem[{{Belkacem} {et~al}\mbox{.}(2008){Belkacem}, {Samadi}, {Goupil}, \&
  {Dupret}}]{belkacem2008}
{Belkacem} K., {Samadi} R., {Goupil} M.-J., {Dupret} M.-A., 2008, \aap, 478,
  163

\bibitem[{{Belkacem} {et~al}\mbox{.}(2009{\natexlab{a}}){Belkacem}, {Samadi},
  {Goupil}, {Dupret}, {Brun}, \& {Baudin}}]{belkacem2009}
{Belkacem} K., {Samadi} R., {Goupil} M.~J., {Dupret} M.~A., {Brun} A.~S.,
  {Baudin} F., 2009{\natexlab{a}}, \aap, 494, 191

\bibitem[{{Belkacem} {et~al}\mbox{.}(2009{\natexlab{b}}){Belkacem}, {Samadi},
  {Goupil}, {Lef{\`e}vre}, {Baudin}, {Deheuvels}, {Dupret}, {Appourchaux},
  {Scuflaire}, {Auvergne}, {Catala}, {Michel}, {Miglio}, {Montalban}, {Thoul},
  {Talon}, {Baglin}, \& {Noels}}]{belkacem2009science}
{Belkacem} K. {et~al.}, 2009{\natexlab{b}}, Science, 324, 1540

\bibitem[{{Blomme} {et~al}\mbox{.}(2011){Blomme}, {Mahy}, {Catala}, {Cuypers},
  {Gosset}, {Godart}, {Montalban}, {Ventura}, {Rauw}, {Morel}, {Degroote},
  {Aerts}, {Noels}, {Michel}, {Baudin}, {Baglin}, {Auvergne}, \&
  {Samadi}}]{blomme2011}
{Blomme} R. {et~al.}, 2011, \aap, 533, A4

\bibitem[{{Briquet} {et~al}\mbox{.}(2007){Briquet}, {Morel}, {Thoul},
  {Scuflaire}, {Miglio}, {Montalb{\'a}n}, {Dupret}, \& {Aerts}}]{briquet2007}
{Briquet} M., {Morel} T., {Thoul} A., {Scuflaire} R., {Miglio} A.,
  {Montalb{\'a}n} J., {Dupret} M.-A., {Aerts} C., 2007, \mnras, 381, 1482

\bibitem[{{Brott} {et~al}\mbox{.}(2011){Brott}, {de Mink}, {Cantiello},
  {Langer}, {de Koter}, {Evans}, {Hunter}, {Trundle}, \& {Vink}}]{brott2011}
{Brott} I. {et~al.}, 2011, \aap, 530, A115

\bibitem[{{Browning} {et~al}\mbox{.}(2004){Browning}, {Brun}, \&
  {Toomre}}]{browning2004}
{Browning} M.~K., {Brun} A.~S., {Toomre} J., 2004, \apj, 601, 512

\bibitem[{Cantiello {et~al}\mbox{.}(2009)Cantiello, Langer, Brott, de~Koter,
  Shore, Vink, Voegler, Lennon, \& Yoon}]{cantiello2009}
Cantiello M. {et~al.}, 2009, Astronomy and Astrophysics, 499, 279

\bibitem[{{Christensen-Dalsgaard}(2002)}]{dalsgaard2002}
{Christensen-Dalsgaard} J., 2002, Reviews of Modern Physics, 74, 1073

\bibitem[{{Christensen-Dalsgaard} {et~al}\mbox{.}(2011){Christensen-Dalsgaard},
  {Monteiro}, {Rempel}, \& {Thompson}}]{dalsgaard2011}
{Christensen-Dalsgaard} J., {Monteiro} M.~J.~P.~F.~G., {Rempel} M., {Thompson}
  M.~J., 2011, \mnras, 414, 1158

\bibitem[{{De Cat}(2007)}]{de-cat2007}
{De Cat} P., 2007, Communications in Asteroseismology, 150, 167

\bibitem[{{de Jager} {et~al}\mbox{.}(1988){de Jager}, {Nieuwenhuijzen}, \& {van
  der Hucht}}]{de-jager1988}
{de Jager} C., {Nieuwenhuijzen} H., {van der Hucht} K.~A., 1988, \aaps, 72, 259

\bibitem[{{Dupret} {et~al}\mbox{.}(2009){Dupret}, {Belkacem}, {Samadi},
  {Montalban}, {Moreira}, {Miglio}, {Godart}, {Ventura}, {Ludwig},
  {Grigahc{\`e}ne}, {Goupil}, {Noels}, \& {Caffau}}]{dupret2009}
{Dupret} M.-A. {et~al.}, 2009, \aap, 506, 57

\bibitem[{{Dziembowski}(1977)}]{dziembowski1977}
{Dziembowski} W., 1977, \actaa, 27, 203

\bibitem[{{Garcia Lopez} \& {Spruit}(1991)}]{spruit1991}
{Garcia Lopez} R.~J., {Spruit} H.~C., 1991, \apj, 377, 268

\bibitem[{{Gizon} {et~al}\mbox{.}(2010){Gizon}, {Birch}, \&
  {Spruit}}]{gizon2010}
{Gizon} L., {Birch} A.~C., {Spruit} H.~C., 2010, \araa, 48, 289

\bibitem[{{Goldreich} \& {Kumar}(1990)}]{goldreich1990}
{Goldreich} P., {Kumar} P., 1990, \apj, 363, 694

\bibitem[{{Grevesse} \& {Sauval}(1998)}]{gs98}
{Grevesse} N., {Sauval} A.~J., 1998, \ssr, 85, 161

\bibitem[{{Koch} {et~al}\mbox{.}(2010){Koch}, {Borucki}, {Basri}, {Batalha},
  {Brown}, {Caldwell}, {Christensen-Dalsgaard}, {Cochran}, {DeVore}, {Dunham},
  {Gautier}, {Geary}, {Gilliland}, {Gould}, {Jenkins}, {Kondo}, {Latham},
  {Lissauer}, {Marcy}, {Monet}, {Sasselov}, {Boss}, {Brownlee}, {Caldwell},
  {Dupree}, {Howell}, {Kjeldsen}, {Meibom}, {Morrison}, {Owen}, {Reitsema},
  {Tarter}, {Bryson}, {Dotson}, {Gazis}, {Haas}, {Kolodziejczak}, {Rowe}, {Van
  Cleve}, {Allen}, {Chandrasekaran}, {Clarke}, {Li}, {Quintana}, {Tenenbaum},
  {Twicken}, \& {Wu}}]{kepler}
{Koch} D.~G. {et~al.}, 2010, \apjl, 713, L79

\bibitem[{{Kumar} \& {Quataert}(1997)}]{kumar1997}
{Kumar} P., {Quataert} E.~J., 1997, \apjl, 475, L143

\bibitem[{{Kumar} {et~al}\mbox{.}(1996){Kumar}, {Quataert}, \&
  {Bahcall}}]{kumar1996}
{Kumar} P., {Quataert} E.~J., {Bahcall} J.~N., 1996, \apjl, 458, L83

\bibitem[{{Kumar} {et~al}\mbox{.}(1999){Kumar}, {Talon}, \& {Zahn}}]{kumar1999}
{Kumar} P., {Talon} S., {Zahn} J., 1999, \apj, 520, 859

\bibitem[{{Lebreton} \& {Goupil}(2012)}]{lebreton2012}
{Lebreton} Y., {Goupil} M., 2012, ArXiv e-prints

\bibitem[{{Lecoanet} \& {Quataert}(2012)}]{lecoanet2012}
{Lecoanet} D., {Quataert} E., 2012, ArXiv e-prints

\bibitem[{{Paxton} {et~al}\mbox{.}(2011){Paxton}, {Bildsten}, {Dotter},
  {Herwig}, {Lesaffre}, \& {Timmes}}]{mesa2011}
{Paxton} B., {Bildsten} L., {Dotter} A., {Herwig} F., {Lesaffre} P., {Timmes}
  F., 2011, \apjs, 192, 3

\bibitem[{{Press}(1981)}]{press1981}
{Press} W.~H., 1981, \apj, 245, 286

\bibitem[{{Rogers} \& {Glatzmaier}(2005)}]{rogers2005}
{Rogers} T.~M., {Glatzmaier} G.~A., 2005, \mnras, 364, 1135

\bibitem[{{Rogers} \& {Glatzmaier}(2006)}]{rogers2006}
{Rogers} T.~M., {Glatzmaier} G.~A., 2006, \apj, 653, 756

\bibitem[{{Rogers} {et~al}\mbox{.}(2006){Rogers}, {Glatzmaier}, \&
  {Jones}}]{rogers2006a}
{Rogers} T.~M., {Glatzmaier} G.~A., {Jones} C.~A., 2006, \apj, 653, 765

\bibitem[{{Rogers} {et~al}\mbox{.}(2012){Rogers}, {Lin}, \& {Lau}}]{rogers2012}
{Rogers} T.~M., {Lin} D.~N.~C., {Lau} H.~H.~B., 2012, \apjl, 758, L6

\bibitem[{{Samadi} {et~al}\mbox{.}(2010){Samadi}, {Belkacem}, {Goupil},
  {Dupret}, {Brun}, \& {Noels}}]{samadi2010}
{Samadi} R., {Belkacem} K., {Goupil} M.~J., {Dupret} M.-A., {Brun} A.~S.,
  {Noels} A., 2010, \apss, 328, 253

\bibitem[{{Shiode} {et~al}\mbox{.}(2012){Shiode}, {Quataert}, \&
  {Arras}}]{shiode2012}
{Shiode} J.~H., {Quataert} E., {Arras} P., 2012, \mnras, 3116

\bibitem[{{Sim{\'o}n-D{\'{\i}}az} {et~al}\mbox{.}(2011){Sim{\'o}n-D{\'{\i}}az},
  {Castro}, {Herrero}, {Aerts}, {Puls}, \& {Markova}}]{simon-diaz2011}
{Sim{\'o}n-D{\'{\i}}az} S., {Castro} N., {Herrero} A., {Aerts} C., {Puls} J.,
  {Markova} N., 2011, ArXiv e-prints

\bibitem[{{Turner} {et~al}\mbox{.}(2004){Turner}, {Yorke}, {Socrates}, \&
  {Blaes}}]{turner2004}
{Turner} N.~J., {Yorke} H.~W., {Socrates} A., {Blaes} O.~M., 2004, in Revista
  Mexicana de Astronomia y Astrofisica Conference Series, Vol.~22, Revista
  Mexicana de Astronomia y Astrofisica Conference Series, {Garcia-Segura} G.,
  {Tenorio-Tagle} G., {Franco} J., {Yorke} H.~W., eds., pp. 54--57

\bibitem[{{Unno} {et~al}\mbox{.}(1989){Unno}, {Osaki}, {Ando}, {Saio}, \&
  {Shibahashi}}]{unno}
{Unno} W., {Osaki} Y., {Ando} H., {Saio} H., {Shibahashi} H., 1989, {Nonradial
  oscillations of stars, 2nd ed.} Tokyo: University of Tokyo Press

\bibitem[{{Uytterhoeven} {et~al}\mbox{.}(2011){Uytterhoeven}, {Moya},
  {Grigahc{\`e}ne}, {Guzik}, {Guti{\'e}rrez-Soto}, {Smalley}, {Handler},
  {Balona}, {Niemczura}, {Fox Machado}, {Benatti}, {Chapellier}, {Tkachenko},
  {Szab{\'o}}, {Su{\'a}rez}, {Ripepi}, {Pascual}, {Mathias},
  {Mart{\'{\i}}n-Ru{\'{\i}}z}, {Lehmann}, {Jackiewicz}, {Hekker},
  {Gruberbauer}, {Garc{\'{\i}}a}, {Dumusque}, {D{\'{\i}}az-Fraile}, {Bradley},
  {Antoci}, {Roth}, {Leroy}, {Murphy}, {De Cat}, {Cuypers}, {Kjeldsen},
  {Christensen-Dalsgaard}, {Breger}, {Pigulski}, {Kiss}, {Still}, {Thompson},
  \& {van Cleve}}]{uytterhoeven2011}
{Uytterhoeven} K. {et~al.}, 2011, \aap, 534, A125

\bibitem[{{Vink} {et~al}\mbox{.}(2001){Vink}, {de Koter}, \&
  {Lamers}}]{vink2001}
{Vink} J.~S., {de Koter} A., {Lamers} H.~J.~G.~L.~M., 2001, \aap, 369, 574

\bibitem[{{Winget} \& {Kepler}(2008)}]{winget2008}
{Winget} D.~E., {Kepler} S.~O., 2008, \araa, 46, 157

\bibitem[{{Wolff} {et~al}\mbox{.}(2006){Wolff}, {Strom}, {Dror}, {Lanz}, \&
  {Venn}}]{wolff2006}
{Wolff} S.~C., {Strom} S.~E., {Dror} D., {Lanz} L., {Venn} K., 2006, \aj, 132,
  749

\end{thebibliography}
}

\label{lastpage}

\end{document}